\documentclass[conference]{IEEEtran}
\ifCLASSINFOpdf
\else
\fi
\usepackage{amsmath,dsfont,bbm,epsfig,amssymb,amsfonts,amstext,verbatim,amsopn,cite,subfigure,multirow,multicol,lipsum,xfrac}
\usepackage{mathtools, amsthm}
\usepackage{perpage}
\usepackage{balance}
\usepackage{url}
\usepackage{amsfonts}
\usepackage{epsfig}
\usepackage[font={small}]{caption}
\usepackage{psfrag}	
\usepackage{etoolbox}
\usepackage{algorithmicx}
\usepackage[Algorithm,ruled]{algorithm}
\usepackage{algpseudocode}
\usepackage{pifont}
\usepackage[utf8]{inputenc}
\usepackage[T1]{fontenc}  
\usepackage[nolist]{acronym}
\MakePerPage{footnote}
\usepackage{paralist}
\usepackage{enumitem}
\usepackage[process=auto]{pstool}
\usepackage{bbm}
\usepackage{tikz}
\usetikzlibrary{shapes,arrows}
\usepackage{pgfplots}
\newcommand{\mas}{\mathcal{S}}
\newcommand{\mai}{\mathcal{I}}

\newcommand{\man}{\mathcal{N}}
\newcommand{\maa}{\mathcal{A}}
\newcommand{\mao}{\mathcal{O}}
\newcommand{\mar}{\mathcal{R}}

\newcommand{\E}{\mathbbmss{E}}
\newcommand{\setS}{\mathbbmss{S}}
\newcommand{\loge}{\mathrm{ln}}

\newcommand{\bx}{\boldsymbol{x}}
\newcommand{\by}{\boldsymbol{y}}
\newcommand{\bn}{\boldsymbol{n}}
\newcommand{\bzero}{\boldsymbol{0}}

\newcommand{\mH}{\mathbf{H}}
\newcommand{\mDelta}{\mathbf{\Delta}}
\newcommand{\tmH}{\tilde{\mathbf{H}}}
\newcommand{\mI}{\mathbf{I}}

\newcommand{\mQ}{\mathbf{Q}}
\newcommand{\mJ}{\mathbf{J}}

\newcommand{\bfh}{\mathbf{h}}
\newcommand{\tbfh}{\tilde{\mathbf{h}}}

\newcommand{\abs}[1]{\vert #1 \vert}
\newcommand{\norm}[1]{\Vert #1 \Vert}
\newcommand{\set}[1]{\left\lbrace #1 \right\rbrace}
\newcommand{\tr}[1]{\mathrm{Tr}\left\lbrace #1 \right\rbrace}

\newtheoremstyle{mystyle}
  {}
  {}
  {}
  {}
  {\bfseries}
  {.}
  { }
  {}

\theoremstyle{mystyle}

\newtheorem*{prf}{Proof}
\newtheorem{theorem}{Proposition}
\newtheorem{lemma}{Lemma}
\newtheorem*{remark}{Remark}
\newtheorem*{spec}{Special Case}

\newcommand{\cnormal}{\mathcal{CN}}
\newcommand{\her}{\mathsf{H}}

\def\nsp{\hskip\fontdimen2\font\relax}

\begin{document}
%
\title{Asymptotics of Transmit Antenna Selection:\\Impact of Multiple Receive Antennas\vspace*{-3mm}}

\author{\IEEEauthorblockN{Saba Asaad\IEEEauthorrefmark{1}\IEEEauthorrefmark{2}, Ali Bereyhi\IEEEauthorrefmark{2}, Ralf R. M\"uller\IEEEauthorrefmark{2}, Amir M. Rabiei\IEEEauthorrefmark{1}}
\IEEEauthorblockA{\IEEEauthorrefmark{1}School of Electrical and Computer Engineering, University of Tehran\\
\IEEEauthorrefmark{2}Institute for Digital Communications (IDC), Friedrich-Alexander Universit\"at Erlangen-N\"urnberg (FAU)\\
saba{\_}asaad@ut.ac.ir, ali.bereyhi@fau.de, ralf.r.mueller@fau.de, rabiei@ut.ac.ir\vspace*{-4mm}
\thanks{This work was supported by the German Research Foundation (Deutsche Forschungsgemeinschaft, DFG) under Grant No. MU 3735/2-1.}
}}



%


\IEEEoverridecommandlockouts

\maketitle

\begin{abstract}
$ \ $Consider a fading Gaussian MIMO channel with $N_\mathrm{t}$ transmit and $N_\mathrm{r}$~receive antennas. The transmitter selects~$L_\mathrm{t}$ antennas corresponding to the strongest channels. For this setup, we study the distribution of the input-output mutual information when $N_\mathrm{t}$ grows large. We show that, for any $N_\mathrm{r}$~and~$L_\mathrm{t}$,~the~distribution of the input-output mutual information is accurately approximated by a Gaussian distribution whose mean grows large and whose variance converges to zero. Our analysis depicts that, in the large limit, the gap between the expectation of the mutual information and its corresponding upper bound, derived by applying Jensen's inequality, converges to a constant which only depends on $N_\mathrm{r}$ and $L_\mathrm{t}$. The result extends the scope of channel hardening to the general case of antenna selection with multiple receive and selected transmit antennas. Although the analyses are given for the large-system limit, our numerical investigations indicate the robustness of the approximated distribution even when the number of antennas is not large.
\end{abstract}
\IEEEpeerreviewmaketitle
\section{Introduction}
Massive \ac{mimo} systems have recently received a great deal of interest due to their promise of high performance gains
\cite{marzetta2010noncooperative}. These gains are mainly achieved at the expense of having a tremendous number of antenna elements within a relatively small physical platform.
From practical points of view, moving to millimeter wave spectrum can make this issue conceivable \cite{rappaport2013millimeter}. The growth in the number of antennas, however, increases the \ac{rf} cost significantly. 
Therefore, addressing solutions to alleviate this issue has become a topic of interest. Antenna selection is a possible solution which reduces hardware costs dramatically without significant performance loss \cite{molisch2005capacity, bai2007channel}. 
For \ac{mimo} channels with limited number of antennas, the performance under different measures, such as capacity, \ac{snr} at the receiver, bit error rate, and outage probability has been investigated in the literature \cite{molisch2005capacity}. There are, however, few results which have addressed the large-system analysis of antenna selection \cite{hesami2011limiting, li2014energy}, and the asymptotic behavior for a general \ac{mimo} setup~is~still~unknown.
\subsection*{Asymptotic Channel Hardening}
The asymptotic hardening property of \ac{mimo} fading channels was first studied in \cite{hochwald2004multiple}. The property indicates that in the large-system limit, the distribution of the mutual information between a white Gaussian input and the output of a \ac{mimo} Gaussian fading channel concentrates almost normally around its mean while the variance shrinks rapidly. 
Under single \ac{tas}, the channel hardening was studied initially in \cite{bai2007channel}, where the authors considered a \ac{tas} protocol selecting a single transmit antenna with the strongest channel gain. It was further shown that, under this scheme, the channel hardens at a slower rate compared to the case considered in \cite{hochwald2004multiple}. In \cite{hesami2011limiting}, the limiting behavior of the mutual information in an uplink channel was investigated considering the transmitter to be equipped with a single transmit antenna, and the receiver to select a number of strongest channels. For this scenario, the distribution of the input-output mutual information was approximated with the distribution of the logarithm of a folded normal random variable. In the large-system limit, it was further shown that the variance converges to zero which concluds the asymptotic hardening property for the setup. The asymptotics of \ac{tas} were further studied in \cite{li2014energy} for a downlink scenario with single antenna receiver in which a multi-antenna transmitter selects a number of antennas with strongest channel coefficients. For this scenario, the input-output mutual information was approximated asymptotically, and the hardening property was shown to hold.

In this paper, we generalize the earlier studies by determining a large-system approximation for the input-output mutual information of a Gaussian \ac{mimo} channel, when both the transmitter and receiver are equipped with multiple antennas and the transmitter selects a finite number of transmit antennas. The \ac{tas} protocol, considered here, selects the antennas which observe the strongest channel gains to the receiver. Our results show that in the asymptotic regime, the gap between the expected mutual information and the upper bound derived by applying Jensen's inequality, remains constant in terms of $N_\mathrm{t}$. Using the large-system approximation, we further investigate the asymptotic channel hardening property for this setup. For the special case of single-antenna receiver, our result reduces to the approximation reported in the literature.

\textit{Notation:} Scalars, vectors and matrices~are~represented~with non-bold, bold lower case and bold upper case letters, respectively.~$\mH^{\her}$ indicates the Hermitian of $\mH$, and $\mI_N$ is the $N\times N$ identity matrix. The determinant of $\mH$ and euclidean norm of $\bx$ are denoted by $\abs{\mH}$ and $\norm{\bx}$. $\log\left(\cdot\right)$ and $\loge \left(\cdot\right)$ indicate the binary and natural logarithm, and $\E\set{\cdot}$ is the expectation operator. The beta distribution with the shape parameters $\alpha$ and $\beta$ is denoted by $\mathrm{Beta}(\alpha,\beta)$.
\section{Problem Formulation}
We\nsp consider a Gaussian \ac{mimo} channel~in~which the transmitter and receiver are equipped with~$N_\mathrm{t}$~and~$N_\mathrm{r}$~antennas, respectively. The transmitter selects $L_\mathrm{t}$ transmit antennas based on the information provided through a rate-limited return~channel. 
For this~setup,~we investigate the input-output~mutual~information when the number of transmit antennas grows large.
\subsection{System Model}
\label{sec:sys}
The received signal by the receiver at each time interval is denoted by $\by_{N_\mathrm{r}\times 1}$ and reads
\begin{align}
\by=\sqrt{\rho} \ \mH \bx + \bn, \label{eq:sys-1}
\end{align}
where $\rho$ denotes the average~\ac{snr}~at~each receive~antenna, $\bn_{N_\mathrm{r} \times 1}$ is circularly symmetric zero-mean complex Gaussian noise with unit variance, i.e., $\bn \sim \cnormal (\bzero, \mI)$, $\bx_{N_\mathrm{t} \times 1}$ identifies the transmit signal with the power constraint $\E \bx^\her \bx \leq 1$, and $\mH$ denotes an ${N_\mathrm{r} \times N_\mathrm{t}}$ \ac{iid} unit-variance Rayleigh fading channel.
It is assumed that the \ac{csi} is available only at the receiver side.
\subsection{\ac{tas} Protocol}
\label{sec:tas}
The transmitter, at each time interval, selects the $L_\mathrm{t}$ strong-est channels by employing the \ac{tas} protocol $\mas$. To illustrate the protocol, let $\bfh_j$ denote the $j$th column vector of $\mH$ for $j\in \set{1, \ldots, N_\mathrm{t}}$. Moreover, represent the index set of order statistics from the arranging of vectors $\norm{\bfh_j}^2$ in decreasing order of magnitude by $\left\lbrace w_1, \ldots, w_{N_\mathrm{t}} \right\rbrace$, i.e., 
\begin{align}
\norm{\bfh_{w_1}}^2 \geq \norm{\bfh_{w_2}}^2 \geq \cdots \geq \norm{\bfh_{w_{N_\mathrm{t}}}}^2. \label{eq:sys-4}
\end{align}
At each time interval, the receiver informs the transmitter ab-out the set $\left\lbrace w_1, \ldots, w_{L_\mathrm{t}} \right\rbrace$ through a rate-limited return channel. The transmitter, then, selects the corresponding antennas.
\subsection{Input-Output Mutual Information}
Suppose that independent Gaussian symbols are transmitted on the selected antennas. In this case, 
the mutual information between the input vector $\bx$ and the output $\by$ denoted in \eqref{eq:sys-1}, for a given realization of $\mH$, is written as \cite{telatar1999}
\begin{align}
\mai(\mH; \rho \mQ)\coloneqq \log\abs{\mI_{N_\mathrm{r}}+\rho \mH \mQ \mH^\her} \label{eq:sys-6}
\end{align}
where $\mQ$ is an ${N_\mathrm{t} \times N_\mathrm{t}}$ diagonal matrix with nonzero diagonal entries at the indices $\left\lbrace w_1, \ldots, w_{L_\mathrm{t}} \right\rbrace$ and zero at the rest. When the power is uniformly allocated among the selected antennas, the nonzero entries of $\mQ$ equal to $L_\mathrm{t}^{-1}$, and therefore, the input-output mutual information reduces to
\begin{align}
\mai_{\mas}\coloneqq \mai(\tmH; \frac{\rho}{L_\mathrm{t}} \mI_{L_\mathrm{t}}) =\log\abs{\mI_{N_\mathrm{r}}+\frac{\rho}{L_\mathrm{t}}\tmH \tmH^\her} \label{eq:6} 
\end{align}
where $\tmH$ is an $N_\mathrm{r}\times L_\mathrm{t}$ matrix describing the effective channel between the transmitter and the receiver, and constructed from $\mH$ by collecting the columns which correspond to the selected antennas, i.e., $\tmH=[\tbfh_1, \ldots, \tbfh_{L_\mathrm{t}} ]$, where $\{\tbfh_1, \ldots, \tbfh_{L_\mathrm{t}} \}$ is a permutation of $\{\bfh_{w_1}, \ldots, \bfh_{w_{L_\mathrm{t}}} \}$.
For a given~realization~of~$\mH$, $\mai_{\mas}$ upper bounds achievable transmit rates, since the \ac{csi} is only available at the receiver. Thus, one can define $\mai_{\mas}$ to be the maximum achievable rate under the \ac{tas} protocol $\mas$.

\section{Asymptotic Hardening Property under \ac{tas}}
When the transmitter employs all the transmit antennas with uniform power allocation, the input-output mutual information is determined by letting $\mQ=N_\mathrm{t}^{-1} \mI_{N_\mathrm{t}}$ in \eqref{eq:sys-6}. In this~case,~as $N_\mathrm{t}$ grows large with a fixed $N_\mathrm{r}$, $N_\mathrm{t}^{-1}\mH\mH^\her$ converges~to~$\mI_{N_\mathrm{r}}$ due to the law of large numbers, and thus, the mutual information converges to $N_\mathrm{r} \log\left(1+\rho\right)$ in large limits. This property is known as the ``asymptotic hardening'' property and has been rigorously justified for \ac{iid} Rayleigh fading channels when the number of antennas at one side grows large \cite{hochwald2004multiple}.
\subsection{Channel Hardening under \ac{tas}}
Considering the \ac{tas} protocol $\mas$, the channel matrix is a finite collection of order statistics obtained from arranging the magnitude of \ac{iid} channel vectors. The case differs from which considered in \cite{hochwald2004multiple}, and therefore, the asymptotic analyses therein can not be extended. For this case, one can write
\begin{align}
\mai_{\mas}=\sum_{\ell=1}^{L} \log \left( 1+ \frac{\rho}{L_\mathrm{t}} \lambda_\ell \right) \label{eq:eig_val}
\end{align}
where $L\coloneqq\min\set{L_\mathrm{t},N_\mathrm{r}}$, and $\lambda_\ell$ identifies the $\ell$th eigenvalue of $\mJ_{L\times L}$ defined as
 \begin{equation}
 \label{eq:J}
    \mJ=
    \begin{cases}
      \tmH^\her \tmH, & \text{if}\ L=L_\mathrm{t} \\
      \tmH \tmH^\her, & \text{if}\ L=N_\mathrm{r}.
    \end{cases}
  \end{equation}
Consequently, Jensen's inequality \cite{cover2012elements} suggests that
\begin{subequations}
\begin{align}
\mai_{\mas} &\stackrel{\ast}{\leq} L \log \left( 1+ \frac{\rho}{L_\mathrm{t} L} \sum_{\ell=1}^L \lambda_\ell \right) \label{eq:7a} \\
&= L \log \left( 1+ \frac{\rho}{L_\mathrm{t} L} \tr{\mJ}\right) \label{eq:7b}
\end{align}
\end{subequations}
where the equality in $\ast$ holds when $L=1$. Considering the upper bound in \eqref{eq:7b}, $\tr{\mJ}$ is the sum of limited number of order statistics taken from a large arranging set. The sum is known as a trimmed sum in the literature and is shown to converge to a Gaussian random~variable~in~distribution,~when $N_\mathrm{t}$ grows large. This property of trimmed sums justifies the asymptotic hardening property for the upper bound. We approve that the property extends to $\mai_\mas$ as well by approximating the mutual information from the upper bound geometrically. 

\subsection{Main Result}
Proposition \ref{thm:2} approximates $\mai_\mas$ for large $N_{\mathrm{t}}$ with a Gaussian random variables whose mean lies within a fixed~gap~below the mean of the upper bound in \eqref{eq:7b} and whose variance converges to zero as $N_{\mathrm{t}}\uparrow\infty$.
\begin{theorem}
\label{thm:2}
Consider the \ac{tas} protocol $\mas$. For large $N_\mathrm{t}$, $\mai_{\mas}$ is approximated with a Gaussian random variable with mean $\eta$ and variance $\sigma^2$ where
\begin{subequations}
\begin{align}
\eta &= L\left[ \log \left( 1+ \frac{\rho \eta_t}{L_\mathrm{t} L}\right) - \frac{\left(L-1 \right)\rho^2\eta_t^2 \log e}{2M\left( L_\mathrm{t} L + \rho \eta_t \right)^2}  \right] \label{eq:etam} \\
\sigma^2 &= \left[\frac{ \xi L \rho}{L_\mathrm{t}L+\rho \eta_t} \log e \right]^2 \sigma_t^2 \label{eq:sigmam}
\end{align}
\end{subequations}
where $\eta_t$ and $\sigma_t^2$ are determined in Appendix \ref{appendix:b} and read
\begin{subequations}
\begin{align}
\eta_t &= N_\mathrm{r} L_\mathrm{t} \left[ 1 + \mao(\loge \left(\frac{N_\mathrm{t}}{L_\mathrm{t}}\right)) \right] \label{eq:12a}\\
\sigma_t^2 &= N_\mathrm{r} L_\mathrm{t} \left[ N_\mathrm{r}+1 - \mao(\frac{L_\mathrm{t}^{N_\mathrm{r}}}{N_\mathrm{t}^{N_\mathrm{r}}}) \right] \label{eq:12b}.
\end{align}
\end{subequations}
and $\xi$ is given by 
\begin{align}
\xi=1-\frac{ L_\mathrm{t} L \left( L-1 \right)\rho \eta_t }{ M \left( L_\mathrm{t} L + \rho \eta_t \right)^2}.
\end{align}
with $L\coloneqq\min\set{L_\mathrm{t},N_\mathrm{r}}$ and $M\coloneqq\max\set{L_\mathrm{t},N_\mathrm{r}}$.
\end{theorem}
\begin{prf}
The proof is sketched through the large-system analysis in Section~\ref{sec:asym}. The details, however, are left for the extended version of the paper.
\end{prf}

Proposition \ref{thm:2} illustrates the asymptotic hardening property of the \ac{mimo} channel, under the \ac{tas} protocol $\mas$. In fact, as $N_\mathrm{t}$ grows large, $\eta_t$ grows proportionally large, and $\sigma_t^2\ll \eta_t$. Consequently, $\sigma^2 \downarrow 0$ and $\eta$ reads
\begin{align}
\eta \to L \log \left( 1+ \frac{\rho \eta_t}{L_\mathrm{t} L}\right) -  \frac{L \left(L-1 \right)}{2M} \log e. \label{eq:asy_eta}  
\end{align}
Comparing the mean and variance of $\mai_{\mas}$ with the upper bound derived by Jensen's inequality, one observes that
\begin{align}
\lim_{N_\mathrm{t}\uparrow\infty}  \log \left( 1+ \frac{\rho}{L_\mathrm{t} L} \tr{\mJ}\right) - \frac{1}{L} \mai_{\mas}=\frac{L-1}{2M} \log e. \label{eq:main_result_2}
\end{align}
\eqref{eq:main_result_2} states that the gap between $\mai_{\mas}$ and the upper bound given by Jensen's inequality remains fixed asymptotically; the property which indicates that both $\mai_{\mas}$ and the upper bound exhibit a same limiting fluctuation. Noting that $\E\set{\mai_{\mas}}$ grows large proportional to $N_\mathrm{t}$, the upper bound proposed by Jensen's inequality can be considered as a robust measure describing the asymptotics of the input-output mutual information within a constant scalar.
\begin{spec}
Our main result recovers the special case of $N_\mathrm{r}=1$ studied in \cite{li2014energy}. In fact, in this case $L=1$, and thus,~$\xi=1$ which results in same $\eta$ and $\sigma^2$ reported in \cite{li2014energy}. The authors in \cite{li2014energy}, moreover, considered the absolute value of the asymptotic Gaussian random variable in Proposition \ref{thm:2} to be the approximation, in order to avoid approximating negative mutual information. The final expression, however, does not differ from the Gaussian random variable significantly, since probability of $\mai_{\mas}$ being approximated with a negative value is almost zero for large $N_\mathrm{t}$.
\end{spec}
\section{Numerical Results}
\begin{figure}[t]
\hspace*{-.75cm}  
\resizebox{1.1\linewidth}{!}{
\pstool[width=.35\linewidth]{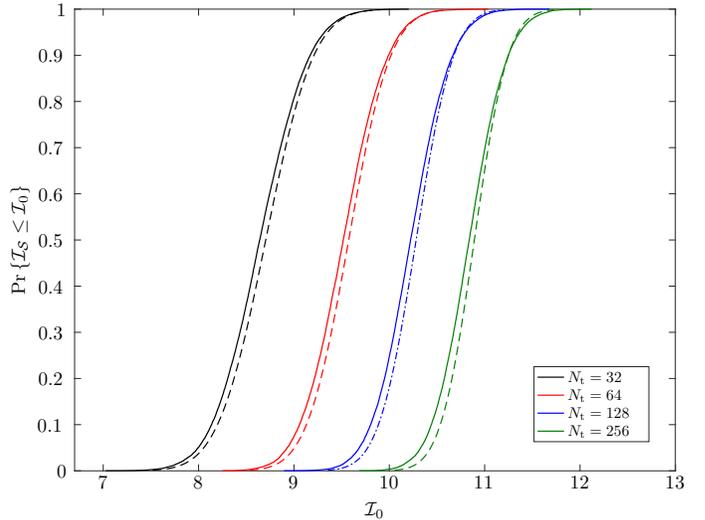}{
\psfrag{Pr}[c][c][0.25]{$\Pr\set{\mai_{\mas}\leq \mai_0}$}
\psfrag{i}[c][c][0.25]{$\mai_0$}
\psfrag{Nt=320AAAB}[l][l][0.2]{$N_\mathrm{t}=32$}
\psfrag{Nt=640AAAB}[l][l][0.2]{$N_\mathrm{t}=64$}
\psfrag{Nt=128AAAB}[l][l][0.2]{$N_\mathrm{t}=128$}
\psfrag{Nt=256AAAB}[l][l][0.2]{$N_\mathrm{t}=256$}

\psfrag{6}[c][c][0.25]{$6$}
\psfrag{7}[c][c][0.25]{$7$}
\psfrag{8}[c][c][0.25]{$8$}
\psfrag{9}[c][c][0.25]{$9$}
\psfrag{10}[c][c][0.25]{$10$}
\psfrag{11}[c][c][0.25]{$11$}
\psfrag{12}[c][c][0.25]{$12$}
\psfrag{13}[c][c][0.25]{$13$}

\psfrag{0}[r][c][0.25]{$0$}
\psfrag{0.1}[r][c][0.25]{$0.1$}
\psfrag{0.2}[r][c][0.25]{$0.2$}
\psfrag{0.3}[r][c][0.25]{$0.3$}
\psfrag{0.4}[r][c][0.25]{$0.4$}
\psfrag{0.5}[r][c][0.25]{$0.5$}
\psfrag{0.6}[r][c][0.25]{$0.6$}
\psfrag{0.7}[r][c][0.25]{$0.7$}
\psfrag{0.8}[r][c][0.25]{$0.8$}
\psfrag{0.9}[r][c][0.25]{$0.9$}
\psfrag{1}[r][c][0.25]{$1$}

}}
\caption{Comparison of empirical cumulative distribution of $\mai_{\mas}$ and approximated distribution given by Proposition~\ref{thm:2} for various number of transmit antennas. The solid and dashed lines~indicate the approximated and empirical distribution, respectively. \ac{snr} is set to be $\rho=0$ dB, $N_\mathrm{r}=8$ and $L_\mathrm{t}=16$.}
\label{fig:1}
\end{figure}

For sake of comparison, the empirical cumulative distributions, and the corresponding approximations are demonstrated in Fig.~\ref{fig:1} for various number of transmit antennas. The empirical distributions are obtained using $20000$ channel realizations. Through our simulations, we assume $L_\mathrm{t}=16$, $N_\mathrm{r}=8$ and $\rho=0$~dB. As Fig.~\ref{fig:1} illustrates, the approximations, given by Proposition \ref{thm:2}, meet the empirical distributions even within a finite number of antennas. In fact, although our analyses considered the system in the large limit, the simulations show the validity of the results even in non-asymptotic scenarios. Proposition \ref{thm:2}, enables us to accurately approximate diverse performance measures on fading channels, such as ergodic~and outage capacity which we briefly address in the sequel.
\subsection{Ergodic Capacity}
\begin{figure}[t]
\hspace*{-.75cm}  
\resizebox{1.1\linewidth}{!}{
\pstool[width=.35\linewidth]{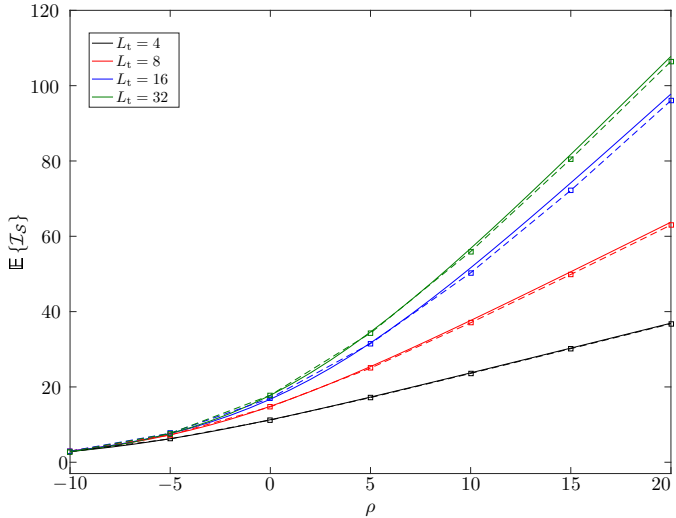}{
\psfrag{EI}[c][c][0.25]{$\E\set{\mai_{\mas}}$}
\psfrag{snr}[c][c][0.25]{$\rho$}
\psfrag{Lt=40AAA}[l][l][0.2]{$L_\mathrm{t}=4$}
\psfrag{Lt=80AAA}[l][l][0.2]{$L_\mathrm{t}=8$}
\psfrag{Lt=16AAA}[l][l][0.2]{$L_\mathrm{t}=16$}
\psfrag{Lt=32AAA}[l][l][0.2]{$L_\mathrm{t}=32$}


\psfrag{5}[c][c][0.25]{\vspace*{5mm}$5$}
\psfrag{10}[c][c][0.25]{$10$}
\psfrag{15}[c][c][0.25]{$15$}
\psfrag{20}[c][c][0.25]{$20$}
\psfrag{25}[c][c][0.25]{$25$}
\psfrag{-10}[c][c][0.25]{$-10$}
\psfrag{-5}[c][c][0.25]{$-5$}
\psfrag{0}[c][c][0.25]{$0$}

\psfrag{20}[r][c][0.25]{$20$}
\psfrag{40}[r][c][0.25]{$40$}
\psfrag{60}[r][c][0.25]{$60$}
\psfrag{80}[r][c][0.25]{$80$}
\psfrag{100}[r][c][0.25]{$100$}
\psfrag{120}[r][c][0.25]{$120$}

}}
\caption{Ergodic capacity as a function of \ac{snr}. The solid lines indicate the approximated ergodic capacity given by Proposition~\ref{thm:2}, and the dashed lines are plotted via numerical simulations. The number of transmit and receive antennas are set to be $N_\mathrm{t}=128$ and $N_\mathrm{r}=16$.}
\label{fig:2}
\end{figure}

For the setup illustrated in Section \ref{sec:sys}, the ergodic capacity is defined as the maximum average achievable transmission rate and given by taking the expectation of the input-output mutual information $\mai_{\mas}$. Using Proposition \ref{thm:2}, the ergodic capacity of the channel $\mH$, under the \ac{tas} protocol $\mas$, is approximated by $\eta$. Fig.~\ref{fig:2} shows the ergodic capacity as a function of average receive \ac{snr} per antenna for various number of selected antennas, when the transmitter and receiver are equipped with $N_\mathrm{t}=128$ and $N_\mathrm{r}=16$ antennas,~respectively. The numerical results show that, for the given range of \ac{snr}s, the approximation tracks the simulation results with maximum of approximately $2\%$ deviation.
\subsection{Outage Capacity}
\begin{figure}[t]
\hspace*{-.75cm}  
\resizebox{1.1\linewidth}{!}{
\pstool[width=.35\linewidth]{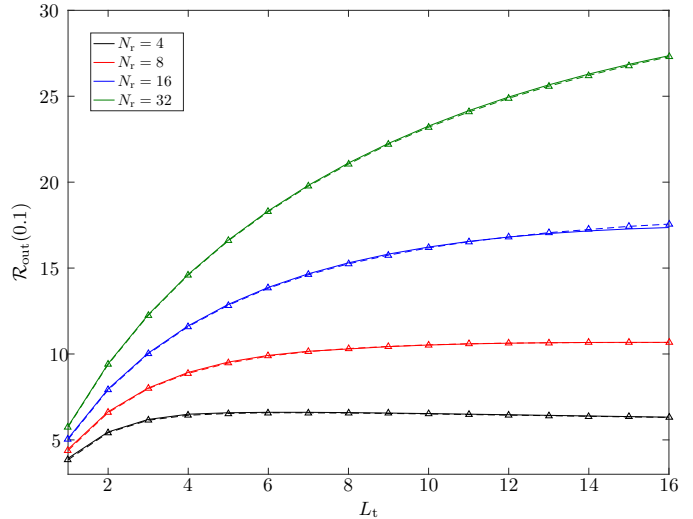}{
\psfrag{Out}[c][c][0.25]{$\mathcal{R}_{\mathrm{out}}(0.1)$}
\psfrag{Lt}[c][c][0.25]{$L_\mathrm{t}$}
\psfrag{Nr=40AAA}[l][l][0.2]{$N_\mathrm{r}=4$}
\psfrag{Nr=80AAA}[l][l][0.2]{$N_\mathrm{r}=8$}
\psfrag{Nr=16AAA}[l][l][0.2]{$N_\mathrm{r}=16$}
\psfrag{Nr=32AAA}[l][l][0.2]{$N_\mathrm{r}=32$}

\psfrag{5}[r][c][0.25]{$5$}
\psfrag{10}[c][c][0.25]{$10$}
\psfrag{15}[r][c][0.25]{$15$}
\psfrag{20}[r][c][0.25]{$20$}
\psfrag{25}[r][c][0.25]{$25$}
\psfrag{30}[r][c][0.25]{$30$}

\psfrag{2}[c][c][0.25]{$2$}
\psfrag{4}[c][c][0.25]{$4$}
\psfrag{6}[c][c][0.25]{$6$}
\psfrag{8}[c][c][0.25]{$8$}
\psfrag{10}[c][c][0.25]{$10$}
\psfrag{12}[c][c][0.25]{$12$}
\psfrag{14}[c][c][0.25]{$14$}
\psfrag{16}[c][c][0.25]{$16$}

}}
\caption{$10\%$ outage capacity in terms of number of selected antennas. The solid and dashed lines respectively denote the approximations, and the numerical simulations for $N_\mathrm{t}=128$ at $\rho=0$ dB.}
\label{fig:3}
\end{figure}

In slow fading scenarios, where the channel does not fluctuate significantly within the transmission interval, the ergodic capacity cannot describe the real transmission limit on the channel. In this case, one may consider the outage capacity $\mar_{\mathrm{out}} (p_{\mathrm{out}})$ which for a given outage probability $p_{\mathrm{out}}$ reads
\begin{align}
\Pr\set{\mai_{\mas} \leq \mar_{\mathrm{out}} (p_{\mathrm{out}})} = 1-p_{\mathrm{out}}.
\end{align}
Fig. \ref{fig:3} plots the $10\%$ outage capacity, i.e., $p_\mathrm{out}=0.1$, versus the number of selected antennas for different $N_\mathrm{r}$, assuming $N_\mathrm{t}=128$ and $\rho=0$~dB. As the figure illustrates, for~the~given numbers of selected antennas, Proposition \ref{thm:2} meets the numerical simulations with approximately $1.5 \%$ deviation at most.
%
\section{Large-System Analysis}
\label{sec:asym}
In this section, we briefly sketch the proof of Proposition~\ref{thm:2}. Due to the lack of space, we omit the detailed derivations here and give them in the extended version of the manuscript. Our derivations mainly follow two steps:
\begin{enumerate}[label=(\Alph*)]
\item An approximation for the input-output mutual information of a Gaussian \ac{mimo} channel is calculated. Using the approximation, $\mai_\mas$ is given in terms of a trimmed~sum.
\item The asymptotic properties of trimmed sums, as well as random matrices, are employed to determine the statistics of the selected channel in the large-system limit.
\end{enumerate}
\subsection{Approximating the Mutual Information}
\label{sec:approx}
Considering the effective channel $\tmH$, we derive an approximation for the input-output mutual information in terms of the first and second order statistics of $\mJ$ defined in \eqref{eq:J}.
Starting from \eqref{eq:eig_val}, let us define the scalars $i_\ell$ for $\ell \in \set{1, \ldots, L}$ to be
\begin{align}
i_\ell=\log \left( 1+ \frac{\rho}{L_\mathrm{t}} \lambda_\ell \right).
\end{align}
The tuple $(\mu , c)$ then denotes the centroid of the set  
\begin{align}
\maa=\set{ (\lambda_\ell , i_\ell) \ \ \forall \ell \in \set{1, \ldots, L}},
\end{align}
and reads
\begin{subequations}
\begin{align}
c&=\frac{1}{L} \sum_{\ell=1}^L i_\ell = \frac{1}{L} \mai_{\mas}, \\
\mu&=\frac{1}{L} \sum_{\ell=1}^L \lambda_\ell = \frac{1}{L} \tr{\mJ}.
\end{align}
\end{subequations}
Therefore, the upper bound in \eqref{eq:7b} reduces to
\begin{align}
c &\leq  \log \left( 1+ \frac{\rho}{L_\mathrm{t}} \mu \right). \label{eq:upper}
\end{align}
\eqref{eq:upper} suggests a geometric approximation for $c$~which~we~intu-itively illustrate here.

\begin{figure}[t]
\hspace*{-.8cm}  
\resizebox{1.1\linewidth}{!}{
\pstool[width=.35\linewidth]{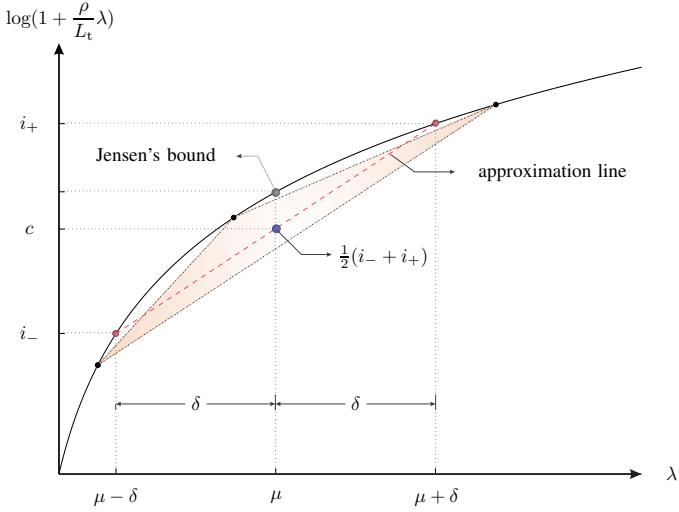}{

\psfrag{my4}[c][c][0.25]{$i_+$}
\psfrag{my2}[c][c][0.25]{$c$}
\psfrag{lam}[c][c][0.25]{$\lambda$}
\psfrag{log}[c][c][0.25]{$\log(1+\dfrac{\rho}{L_\mathrm{t}} \lambda)$}
\psfrag{my1}[c][c][0.25]{$i_-$}
\psfrag{mu}[c][c][0.25]{$\mu$}
\psfrag{vu}[c][c][0.25]{$\mu-\delta$}
\psfrag{xu}[c][c][0.25]{$\mu+\delta$}
\psfrag{approxm}[c][c][0.25]{$\frac{1}{2}(i_- + i_+)$}
\psfrag{approx}[c][c][0.25]{approximation line}
\psfrag{jensen}[c][c][0.25]{Jensen's bound}
\psfrag{3 in}[c][c][0.25]{$\delta$}
}}
\caption{Geometric approximation of $\mai(\mas)$. By choosing $\delta$ properly, the approximation line at $\lambda=\mu$ takes a value close to the centroid of the polytope.\vspace*{-3mm}}
\label{fig:4}
\end{figure}

Consider the mutual information curve $\log \left(1+\rho L_\mathrm{t}^{-1} x\right)$. By deviating from $\mu$ on the curve with a certain step size $\delta$, the tuples $(\mu-\delta, i_-)$ and $(\mu+\delta, i_+)$ are obtained where
\begin{subequations}
\begin{align}
i_- &=\log \left( 1+ \frac{\rho}{L_\mathrm{t}} (\mu-\delta) \right) \label{eq:i_-}, \\
i_+ &=\log \left( 1+ \frac{\rho}{L_\mathrm{t}} (\mu+\delta) \right) \label{eq:i_+}.
\end{align}
\end{subequations}
Due to continuity and concavity of the mutual information curve, the centroid of the line connecting $(\mu-\delta, i_-)$ and $(\mu+\delta, i_+)$ lies under $(\mu,c)$ for some choices of $\delta$. The point, moreover, can be arbitrarily close to $(\mu,c)$, if $\delta$ is set properly; see Fig.~\ref{fig:4}. 
As the result, $\mai_{\mas}$ can be approximated as
\begin{align}
{\mai}_{\mas} &\approx  \frac{L}{2} \left( i_- + i_- \right). \label{eq:apx_mai}
\end{align}
Using the polynomial expansion, \eqref{eq:apx_mai} is written as
\begin{align}
{\mai}_{\mas}= L \left[ \log \left( 1+ \frac{\rho}{L_\mathrm{t}} \mu \right) -\frac{\kappa^2 \delta^2}{2}\log e \right] + \mao(\kappa^4 \delta^4) \label{eq:apprx_i}
\end{align} 
where $\kappa$ is defined to be
\begin{align}
\kappa\coloneqq\frac{\rho}{L_\mathrm{t} + \rho \mu}. \label{eq:kappa}
\end{align}
At this point, we need to determine $\delta$. To do so, we note that the input-output mutual information in \eqref{eq:6}, can be expanded in an alternative way. Define $\mDelta$ to be an $L\times L$ matrix, such~that $\mJ = \mu \mI_L + \mDelta$. Consequently, one can write
\begin{align}
\mai_{\mas} &= L \log \left( 1+ \frac{\rho}{L_\mathrm{t}} \mu \right) + \log \abs{\mI_L + \kappa \mDelta} \label{eq:log}
\end{align} 
where $\kappa$ is defined in \eqref{eq:kappa}. We expand the second term in the \ac{rhs} of \eqref{eq:log} by evaluating~a polynomial expansion for the determinant of a perturbed~identity matrix.
\begin{lemma}
\label{lem:1}
For $\kappa$ in a small vicinity of zero, $\abs{\mI + \kappa \mDelta}$ reads
\begin{align}
\hspace*{-2.8mm}\abs{\mI \hspace*{-1mm}+\hspace*{-1mm} \kappa \mDelta} \hspace*{-1mm}=\hspace*{-1mm} 1 \hspace*{-1mm}+\hspace*{-1mm} \kappa \tr{\mDelta} \hspace*{-1mm}+\hspace*{-1mm} \frac{\kappa^2}{2} \hspace*{-1mm}\left[ \hspace*{-.5mm}\tr{\mDelta}^2 \hspace*{-1mm}-\hspace*{-1mm} \tr{\mDelta^2} \hspace*{-.5mm}\right] \hspace*{-1.3mm}+\hspace*{-.8mm} \mao(\kappa^3).
\end{align}
\end{lemma}
\begin{prf}
Starting from the identity $\abs{e^{\kappa\mDelta}}=e^{\kappa\tr{\mDelta}}$, the proof is concluded after some lines of derivations.
\end{prf}
Using Lemma~\ref{lem:1}, \eqref{eq:log} reduces to
\begin{align}
\mai_{\mas} &= L  \log \left( 1+ \frac{\rho \mu}{L_\mathrm{t}}\right) +\log \left(1-\frac{\kappa^2}{2} \tr{\mDelta^2}+\mao(\kappa^3)\right) \nonumber  \\
&\stackrel{\star}{=} L \log \left( 1+ \frac{\rho\mu}{L_\mathrm{t}} \right) - \frac{\kappa^2}{2} \tr{\mDelta^2} \log e + \mao(\kappa^3) \label{eq:mai}
\end{align}
where $\star$ follows from taking the assumption that $\kappa \downarrow 0$ as $N_\mathrm{t}$ grows large. We later show that the assumption holds, since $\mu$ grows proportional to $N_\mathrm{t}$. By letting the \ac{rhs} of \eqref{eq:apprx_i} and \eqref{eq:mai} to be equal, $\delta$ is determined as
\begin{align}
\delta^2=\frac{\tr{\mDelta^2}}{L } = \frac{\tr{\mJ^2}}{L} - \frac{\tr{\mJ}^2}{L^2  }. \label{eq:del2}
\end{align}
Substituting \eqref{eq:del2} in \eqref{eq:apx_mai}, we conclude the following lemma.
\begin{lemma}
\label{thm:1}
The input-output mutual information for any realization of $\mH$ is approximated as
\begin{align}
\hspace{-1mm}\mai_{\mas} \hspace{-.7mm} \approx \hspace{-.7mm} \frac{L}{2} \left[ \log\left(1\hspace{-.7mm}+\hspace{-.7mm}\frac{\rho\left(\mu\hspace{-.7mm}-\hspace{-.7mm}\delta\right)}{L_\mathrm{t}}\right) \hspace{-.7mm} + \hspace{-.7mm} \log\left(1\hspace{-.7mm}+\hspace{-.7mm}\frac{\rho\left(\mu\hspace{-.7mm}+\hspace{-.7mm}\delta\right)}{L_\mathrm{t}}\right) \vphantom{aa} \right] 
\end{align}
where $\mu \coloneqq {\tr{\mJ}}/{L}$ and $\delta$ is given as in \eqref{eq:del2}.
\end{lemma}
\subsection{Asymptotics of $\mai_{\mas}$}
\label{sec:main}
The approximation proposed in Lemma \ref{thm:1} enables us to investigate the input-output mutual information in the large-system limit. Consider the \ac{tas} protocol $\mas$. As it is indicated in \eqref{eq:sys-4}, ${\norm{\bfh_{w_\ell}}^2}$, for $\ell \in \set{1, \ldots, N_\mathrm{t}}$, are order statistics in decreasing order; therefore, $\tr{\mJ} = \sum_{\ell=1}^{L_\mathrm{t}} \norm{\bfh_{w_\ell}}^2$ is the sum of $L_\mathrm{t}$ first order statistics. In the context of order statistics, this sum is known as a trimmed sum, and shown to converge to a Gaussian random variable in distribution, when the size of arranging set, i.e., $N_\mathrm{t}$, tends to infinity \cite{arnold1992first}. In \cite{stigler1973asymptotic}, the author determined the mean and variance of the asymptotic trimmed sum in terms of the distribution of the arranged random variables. Although the analytic result in \cite{stigler1973asymptotic} is determined in the asymptotic regime, the numerical results show that the distribution of a trimmed sum converges relatively fast to the limiting distribution. Therefore, the asymptotic result provides an efficient and robust approximation of the exact behavior of~a~large~scale,~but~not~infinite~length,~trimmed~sum~\cite{arnold1992first}.

Considering the above discussion, for large $N_\mathrm{t}$, $\tr{\mJ}$ can be approximated with the asymptotic distribution given in \cite{stigler1973asymptotic}, when the distribution of the arranged variables, i.e., $\norm{\bfh_{w_\ell}}^2$, is set to be chi-square with $2N_\mathrm{r}$ degrees of freedom. In this case, by keeping $L_\mathrm{t}$ fixed, the accuracy of the approximation increases, when $N_\mathrm{t}$ grows larger. Using the main theorem of \cite{stigler1973asymptotic}, $\tr{\mJ}$ is approximated with $t$, where $t\sim\man(\eta_t,\sigma_t^2)$. The exact values of $\eta_t$ and $\sigma_t^2$ are determined in Appendix \ref{appendix:b}. Using the large-system approximation, it is shown that the mean and variance for large $N_\mathrm{t}$ are as in \eqref{eq:12a} and \eqref{eq:12b}.

\begin{remark}
As $\tr{\mJ}>0$, the approximation of $\tr{\mJ}$ with a Gaussian random variable clearly fails for some realizations of $\mJ$, since $t$ can take negative values. This is a direct result of the fact that $t$ is only a large-system approximation of $\tr{\mJ}$. As $N_\mathrm{t}$ grows large, $\Pr\set{t <0}$ converges to zero, and therefore, the approximation becomes more accurate.
\end{remark}
As we consider $N_\mathrm{t}$ to be significantly large, one can see that $\kappa \ll 1$ under the \ac{tas} protocol $\mas$. Substituting the large-system approximation of $\tr{\mJ}$ in Lemma \ref{thm:1}, $\mu$ is considered to be normally distributed around $L^{-1} \eta_t$ with variance $L^{-2} \sigma_t^2$. Therefore, using the polynomial expansion, the approximation in Proposition \ref{thm:1} reduces to
\begin{align}
\mai_{\mas}\approx  {L} \log \left( 1+ \frac{\rho}{L_\mathrm{t}L} t  \right)- {\frac{L}{2}} \kappa^2 \delta^2 \log e. \label{eq:final_mai}
\end{align}
The first term in the \ac{rhs} of \eqref{eq:final_mai} is further expanded as
\begin{align}
\log \left( 1\hspace*{-1mm}+ \hspace*{-1mm}\frac{\rho}{L_\mathrm{t}L} t \right) \hspace*{-1.2mm}\stackrel{\dagger}{=}\hspace*{-1mm} \log \left( 1\hspace*{-1mm}+\hspace*{-1mm} \frac{\rho \eta_t}{L_\mathrm{t} L}  \right) \hspace*{-1mm}+\hspace*{-1mm} \frac{\rho (t-\eta_t)}{L_\mathrm{t} L+\rho \eta_t} \log e \hspace*{-1mm}+\hspace*{-1mm} \epsilon_{N_\mathrm{t}} \label{eq:final_mai_3}
\end{align}
where $\epsilon_{N_\mathrm{t}}$ converges to zero as $N_\mathrm{t}$ tends to infinity. $\dagger$ comes from the polynomial expansion of $\log(1+x)$ at $x=0$, in which the higher order terms are dropped. 
Using the same argument as for \eqref{eq:final_mai}-\eqref{eq:final_mai_3}, it is shown that, in the large limit,~$\kappa^2$~reads
\begin{align}
\kappa^2&=\frac{L^2 \rho^2}{\left( L_\mathrm{t} L + \rho \eta_t \right)^2} \left[ 1- 2 \frac{\rho (t-\eta_t)}{ L_\mathrm{t} L+ \rho \eta_t} \right] + \epsilon_{N_\mathrm{t}}. \label{eq:final_kappa_2} 
\end{align}
Although $\delta$ in general finds a complicated distribution, in the large limit, it could be approximated in a straightforward form in terms of $t$. To show that, consider $\mJ$ defined in \eqref{eq:J}. Thus,
\begin{align}
\hspace*{-3mm}\tr{\mJ^2}\hspace*{-.7mm}=\hspace*{-.7mm}
\sum_{\ell=1}^{L_\mathrm{t}} \left[ \norm{\bfh_{w_\ell}}^4 \hspace*{-.7mm}+\hspace*{-.7mm} \norm{\bfh_{w_\ell}}^2 \sum_{\substack{k=1 \\ k\neq \ell}}^{L_\mathrm{t}} \norm{\bfh_{w_k}}^2 \cos^2 \theta_{\ell,k} \right] \label{eq:30b}
\end{align}
where $\theta_{\ell,k}$ denotes the Hermitian angle between $\bfh_\ell$ and $\bfh_k$ and is defined as 
\begin{align}
\theta_{\ell,k} = \cos^{-1} \frac{\abs{\bfh_{w_\ell}^\her \bfh_{w_k}}}{\norm{\bfh_{w_\ell}} \norm{\bfh_{w_k}}}.
\end{align}
By the same approach as the one taken for approximating the distribution of $\tr{\mJ}$, the trace of any principle submatrix of $\mJ$ can be determined in the large limit. In fact, by using the main theorem of \cite{stigler1973asymptotic}, it is shown that the sum of any subset $\setS$ of order statistics $\norm{\bfh_{w_\ell}}^2$ is approximately normally distributed around $\abs{\setS}L_\mathrm{t}^{-1} \eta_t$ with a variance whose tends to zero as $N_\mathrm{t}$ grows large, where $\abs{\setS}\leq L_\mathrm{t}$ indicates the size of $\setS$. Consequently, $\norm{\bfh_{w_\ell}}^2=L_\mathrm{t}^{-1}\tr{\mJ}+\alpha_\ell$ for any $\ell\in\set{1, \ldots, L_\mathrm{t}}$ where $\alpha_\ell$ is a zero-mean random variable with variance~converging to zero. Substituting in \eqref{eq:30b}, 
\begin{align}
\tr{\mJ^2}=\frac{\tr{\mJ}^2}{L_\mathrm{t}}  + \frac{\tr{\mJ}^2}{L_\mathrm{t}^2} \sum_{\substack{\ell,k=1 \\ k\neq \ell}}^{L_\mathrm{t}} \cos^2 \theta_{\ell,k} + \epsilon_{N_\mathrm{t}}, \label{eq:tr2}
\end{align}
where $\epsilon_{N_\mathrm{t}}$ tends to zero as $N_\mathrm{t}$ grows large. Considering~the \ac{tas} protocol $\mas$, the ordering in \eqref{eq:sys-4} considers only the magnitude of $\bfh_j$. Therefore, the distribution of $\theta_{\ell,k}$ is same as the distribution of the Hermitian angles between the column vectors of an \ac{iid} complex Gaussian channel. For this case, it has been reported that $\theta_{\ell,k}$, for a given $\ell$, are independent~\cite{muller1999power}. Moreover, the distribution of $\cos^2 \theta_{\ell,k}$ has been shown to be $\mathrm{Beta}(1,N_\mathrm{r}-1)$, see Appendix C of \cite{muller1999power}. Thus, 
\begin{align}
\tr{\mJ^2}= \frac{\tr{\mJ}^2}{L_\mathrm{t}} \left[ 1+ (L_\mathrm{t}-1)\beta \right] + \epsilon_{N_\mathrm{t}}  \label{eq:tr2_final}
\end{align}
where $\beta$ is defined as $\beta\coloneqq{L_\mathrm{t}}^{-1}\sum_{\ell=1}^{L_\mathrm{t}} \tilde{\beta}_\ell$ with
\begin{align}
\tilde{\beta}_\ell\coloneqq\frac{1}{L_\mathrm{t}-1} \sum_{\substack{k=1, k\neq \ell}}^{L_\mathrm{t}} \cos^2 \theta_{\ell,k}. \label{eq:beta_l}
\end{align}
The summand in \ac{rhs} of \eqref{eq:beta_l} is a sequence of independent beta distributed random variables. Therefore, the distribution of $\tilde{\beta}_\ell$ is given by $L_\mathrm{t}-1$ times convolution of $\mathrm{Beta}(1,N_\mathrm{r}-1)$, and then normalizing correspondingly. Using the properties of beta distribution, it is then shown that even for finite $L_\mathrm{t}$ and $N_\mathrm{r}$, $\tilde{\beta}_\ell$ are approximately distributed normally around $N_\mathrm{r}^{-1}$ with a variance significantly smaller than $N_\mathrm{r}^{-1}$. Consequently, $\beta$ is approximately a Gaussian random variable with mean $N_\mathrm{r}^{-1}$. As $\tilde{\beta}_\ell$ are in general dependent, the variance of $\beta$ is not simply written as the sum of the variances; however, one can upper bound the variance by considering the extreme case of full dependency. Therefore, $\tr{\mJ^2}$ is written as
\begin{align}
\tr{\mJ^2} = \left[ \frac{N_\mathrm{r} + L_\mathrm{t} -1}{N_\mathrm{r} L_\mathrm{t}} + \chi \right] \tr{\mJ}^2 + \epsilon_{N_\mathrm{t}},  \label{eq:tr2_final2}
\end{align}
where $\chi$ is a zero-mean random variable with approximately Gaussian distribution whose variance is relatively small compared to $N_\mathrm{r}^{-1} L_\mathrm{t}^{-1} (N_\mathrm{r} + L_\mathrm{t} -1)$. Defining~$M\coloneqq\max\set{L_\mathrm{t},N_\mathrm{r}}$, the scalar $\delta$ in Lemma~\ref{thm:1} reads
\begin{align}
\delta^2&=\left[ \frac{L-1}{M L^2}+\frac{\chi}{L}\right]  \tr{\mJ}^2 + \epsilon_{N_\mathrm{t}} \label{eq:final_delta_1}.
\end{align}
Using the large-system approximation for $\tr{\mJ}$, i.e., $t$, and taking the same steps as \eqref{eq:final_mai_3}, we have
\begin{align}
\delta^2&=\left[ \frac{L-1}{M L^2}+\frac{\chi}{L}\right] \left[ 1+2\frac{t-\eta_t}{\eta_t} \right] \eta_t^2 . \label{eq:final_delta_2}
\end{align}
Finally by substituting \eqref{eq:final_mai_3}, \eqref{eq:final_kappa_2} and \eqref{eq:final_delta_2} in \eqref{eq:final_mai}, and after some lines of derivations Proposition \ref{thm:2} is concluded.


\appendices
\section{Asymptotics of $\tr{\mJ}$}
\label{appendix:b}
In order to find the exact asymptotic characteristics of the random variable $\tr{\mJ}$ in Section \ref{sec:main}, we invoke the result reported in \cite{stigler1973asymptotic}. Using the main theorem of \cite{stigler1973asymptotic}, $\eta_t$ reads
\begin{align}
\eta_t=N_\mathrm{r}\left[ L_\mathrm{t}  + N_\mathrm{t} f_{N_\mathrm{r} +1}(u) \right]
\end{align}
where $f_{N_\mathrm{r}}(\cdot)$ denotes the chi-square probability density function with $2N_\mathrm{r}$ degrees of freedom and mean $N_\mathrm{r}$,
 \begin{equation}
 \label{eq:Je}
    f_{N_\mathrm{r}}(x)= \frac{1}{(N_\mathrm{r}-1)!}
    \begin{cases}
     x^{N_\mathrm{r}-1} e^{-x} , & \text{if}\ x \geq 0 \\
      0, & \text{if}\ x < 0
    \end{cases}
  \end{equation}
and $u$ is the solution of the equation
\begin{align}
\int_u^\infty f_{N_\mathrm{r}}(x) \mathrm{d} x= \frac{L_\mathrm{t}}{N_\mathrm{t}}.
\end{align}
Moreover, $\sigma_t^2$ is determined as
\begin{align}
\sigma_t^2=\left(L_\mathrm{t}u-\eta_t \right)^2 \left(\frac{1}{L_\mathrm{t}}-\frac{1}{N_\mathrm{t}} \right) - \frac{\eta_t^2}{L_\mathrm{t}}+ \Xi_t
\end{align}
where the non-negative scalar $\Xi_t$ is defined as
\begin{align}
\Xi_t= N_\mathrm{r} \left(N_\mathrm{r}+1\right) \left[ L_\mathrm{t}+N_\mathrm{t} f_{N_\mathrm{r}+1}(u)+N_\mathrm{t} f_{N_\mathrm{r}+2}(u) \right].
\end{align}
\bibliography{ref}

\begin{thebibliography}{10}
\providecommand{\url}[1]{#1}
\csname url@samestyle\endcsname
\providecommand{\newblock}{\relax}
\providecommand{\bibinfo}[2]{#2}
\providecommand{\BIBentrySTDinterwordspacing}{\spaceskip=0pt\relax}
\providecommand{\BIBentryALTinterwordstretchfactor}{4}
\providecommand{\BIBentryALTinterwordspacing}{\spaceskip=\fontdimen2\font plus
\BIBentryALTinterwordstretchfactor\fontdimen3\font minus
  \fontdimen4\font\relax}
\providecommand{\BIBforeignlanguage}[2]{{%
\expandafter\ifx\csname l@#1\endcsname\relax
\typeout{** WARNING: IEEEtran.bst: No hyphenation pattern has been}%
\typeout{** loaded for the language `#1'. Using the pattern for}%
\typeout{** the default language instead.}%
\else
\language=\csname l@#1\endcsname
\fi
#2}}
\providecommand{\BIBdecl}{\relax}
\BIBdecl

\bibitem{marzetta2010noncooperative}
T.~L. Marzetta, ``Noncooperative cellular wireless with unlimited numbers of
  base station antennas,'' \emph{IEEE Trans. on Wireless Communications},
  vol.~9, no.~11, pp. 3590--3600, 2010.

\bibitem{rappaport2013millimeter}
T.~S. Rappaport, S.~Sun, R.~Mayzus, H.~Zhao, Y.~Azar, K.~Wang, G.~N. Wong,
  J.~K. Schulz, M.~Samimi, and F.~Gutierrez, ``Millimeter wave mobile
  communications for 5G cellular: It will work!'' \emph{IEEE access}, vol.~1,
  pp. 335--349, 2013.

\bibitem{molisch2005capacity}
A.~F. Molisch, M.~Z. Win, Y.-S. Choi, and J.~H. Winters, ``Capacity of MIMO
  systems with antenna selection,'' \emph{IEEE Trans. on Wireless
  Communications}, vol.~4, no.~4, pp. 1759--1772, 2005.

\bibitem{bai2007channel}
D.~Bai, P.~Mitran, S.~S. Ghassemzadeh, R.~R. Miller, and V.~Tarokh, ``Channel
  hardening and the scheduling gain of antenna selection diversity schemes,''
  in \emph{IEEE Int. Symp. on Inf. Theory},~pp.~1066--1070,~2007.

\bibitem{hesami2011limiting}
P.~Hesami and J.~N. Laneman, ``Limiting behavior of receive antennae
  selection,'' in \emph{45th
  Annual Conference on Information Sciences and Systems (CISS)}, pp. 1--6, 2011.

\bibitem{li2014energy}
H.~Li, L.~Song, and M.~Debbah, ``Energy efficiency of large-scale multiple
  antenna systems with transmit antenna selection,'' \emph{IEEE Trans. on
  Communications}, vol.~62, no.~2, pp. 638--647, 2014.

\bibitem{hochwald2004multiple}
B.~M. Hochwald, T.~L. Marzetta, and V.~Tarokh, ``Multiple-antenna channel
  hardening and its implications for rate feedback and scheduling,'' \emph{IEEE
  Trans. on Inf. Theory}, vol.~50, no.~9, pp. 1893--1909, 2004.

\bibitem{telatar1999}
E.~Telatar, ``Capacity of multi-antenna Gaussian channels,'' \emph{European
Trans. on Telecomm.}, vol.~10, no.~6, pp.~585–595, 1999.

\bibitem{cover2012elements}
T.~M. Cover and J.~A. Thomas, \emph{Elements of Information Theory}, John Wiley \& Sons, 2012.

\bibitem{arnold1992first}
B.~C. Arnold, N.~Balakrishnan, and H.~N. Nagaraja, \emph{A First Course in
  Order Statistics}, Siam, vol.~54, 1992.

\bibitem{stigler1973asymptotic}
S.~M. Stigler, ``The asymptotic distribution of the trimmed mean,'' \emph{The
  Annals of Statistics}, pp. 472--477, 1973.

\bibitem{muller1999power}
R.~R. M{\"u}ller, ``Multiuser receivers for randomly spread signals: Fundamental limits with and without decision-feedback,'' \emph{IEEE
  Trans. on Inf. Theory}, vol.~47, no.~1, pp. 268--283, 2001.

\end{thebibliography}
\bibliographystyle{IEEEtran}

\begin{acronym}
\acro{mimo}[MIMO]{Multiple-Input Multiple-Output}
\acro{csi}[CSI]{Channel State Information}
\acro{rhs}[r.h.s.]{right hand side}
\acro{awgn}[AWGN]{Additive White Gaussian Noise}
\acro{iid}[i.i.d.]{independent and identically distributed}
\acro{ut}[UT]{User Terminal}
\acro{bs}[BS]{Base Station}
\acro{tas}[TAS]{Transmit Antenna Selection}
\acro{snr}[SNR]{Signal-to-Noise Ratio}
\acro{rf}[RF]{Radio Frequency}
\end{acronym}

\end{document}